\documentclass[12pt,a4paper]{article}
\usepackage[cp1251]{inputenc}
\usepackage[english,russian]{babel}
\usepackage{times}
\usepackage{graphics}
\usepackage{graphicx}
\usepackage{float}
\restylefloat{figure}

\begin{document}

\begin{center}
{\Large\bf Experimental investigation of the low molecular weight

\vspace{1mm}

fluoropolymer for the ultracold neutrons storage}

\vspace{0.5cm}

C. D\"using$^{a}$, P. Geltenbort $^{b}$ C. Plonka$^{a}$,  Yu.N.
Pokotilovski$^{c}$\footnote{corresponding author, e-mail:pokot@nf.jinr.ru}

\vspace {0.5cm}

$^{a}$ Institute of Nuclear Chemistry\\

Johannes Gutenberg University Mainz\\

D-55128 Mainz, Germany

\vskip 0.5cm

              $^{b}$ Institut Laue-Langevin,\\
           BP 156, 38042 Cedex 9  Grenoble, France

\vskip 0.5cm

             $^{c}$ Joint Institute for Nuclear Research\\
              141980 Dubna, Moscow region, Russia\\

\vskip 0.7cm

\begin{minipage}{160mm}

 The experimental setup for examining the low-molecular-weight fluoropolymer
 CF$_{3}$(CF$_{2})_{3}$-O-CF$_{2}$-O-(CF$_{2})_{3}$CF$_{3}$,
which is a promising coating material for the walls of storage chambers for ultracold
neutrons, is described.
 The results are detailed.
 The measurement data are interpreted in the model of a multilayer complex
 quantum-mechanical potential of the chamber walls.

\end{minipage}

\end{center}

\vspace{3mm}


\section{Introduction}
 An accurate neutron lifetime value is needed to refine the Standard Model parameters
 (matrix element Vud) and in cosmology and astrophysics (to characterize the process of
 nucleosynthesis in the early Universe; see recent reviews [1, 2]).

 Two methods for measuring the neutron lifetime are known.
 The (historically) first method involves measurements of the number of neutrons in a
 controlled volume of a beam of slow neutrons and the count rate for the products of
 neutron decay from this volume (in recent experiments these products are protons).
 Absolute measurements of the neutron density in the beam and the number of trapped and
 counted protons are thus required.
 The second method involves the storage of ultracold neutrons (UCN) [3] in a material
 or a magnetic trap and measurements of the variation of the number of neutrons
 remaining in the trap with time.

 The neutron lifetimes determined by these two methods in the most accurate experiments
 in the last 20 years differ considerably.
 The results are summarized in Fig. 1, where the data from two beam experiments [4, 5]
 (the results of the latter study were corrected in [6]), five experiments on UCN
 storage in material traps [7-10, 16] (the results from [10] were later corrected in
 [11, 12]), and the first successful experiments on UCN storage in magnetic traps[13-15]
 are presented.
 Comments on the experiments[7, 9] were given in [17, 18].
 The averaged lifetime derived from beam experiments is $\tau_{beam} = 888.0\pm 2.1$\,s,
 while the corresponding value for the UCN storage experiments (after two corrections
 [11, 12]) is $\tau_{UCN} =879.1\pm 0.4$\,s.
 If the latest results [15, 16] are factored in, the discrepancy between the two methods
  exceeds four standard deviations.

 What are the possible reasons behind this discrepancy?
 The most likely cause is the presence of systematic errors in one of the methods
 (or both methods).
 According to [5], the largest uncertainty in beam experiments comes from the
 determination of the flux of cold neutrons.
 Is it possible that a fraction of protons remains undetected in beam experiments, thus
 increasing the estimate of the neutron lifetime and contributing to the discrepancy?
 The greatest challenge in UCN experiments is the monitoring of UCN losses, and
 inaccurate monitoring may result in erroneously low measured values of the neutron
 lifetime.
 More exotic loss scenarios: of neutron-mirror neutron oscillations [19] or the passage
 of neutrons into extra dimensions or a braneworld [20] (neutron passing-through-walls
 experiments) are also considered.

\section{Issues arising in UCN storage in material traps}

 Neutron losses in material traps are caused by the UCN capture and heating in
 collisions with walls and residual gas atoms.
 The accuracy of determining the neutron lifetime depends on these losses.
 In an effort to enhance this accuracy, experimenters try to reduce the losses and find
 the optimum way of extrapolating the measured UCN storage time to the decay constant
 of a free neutron.
 In some experiments [7, 8, 10-12], these corrections (the difference between the
 measured UCN storage time and the extrapolated lifetime of a free neutron) were fairly
 large (i.e., an order of magnitude larger than the stated lifetime error).
 Therefore, the most obvious way toward enhancing the accuracy of neutron lifetime
 measurements in UCN storage experiments and making the extrapolation more reliable is
 to reduce the UCN losses in traps, thus also reducing the correction for losses.

 There is a known anomaly in UCN losses: the measured and predicted loss coefficients
 for most studied promising materials with small neutron capture cross sections (
 Be, C, D$_{2}$O, ice, and solid oxygen) differ greatly.
 The difference between the measured losses and the calculated ones is especially large
 at low temperatures, where the expected loss coefficients for the mentioned materials
 are exceptionally low ($10^{-6}-10^{-7}$).
 The calculations of expected neutron losses are based on the known dynamical properties
  of materials and the measurements of neutron transmission cross sections.

 In a few experiments with pure (with hydrogen and chlorine impurities removed)
 fluoropolymers in liquid[21] and frozen [9, 22] states, the experimental data were in
 reasonable agreement with the theoretical estimates.
 These polymers (e.g., room temperature liquid Fomblin Y 18/8 [23] in [7, 8, 10-12])
 were used as coatings for trap walls.
 UCN losses on the surface of a liquid polymer are generally well understood and are
 largely attributable to the heating of neutrons by thermal vibrations of atoms of the
 reflecting wall and quasielastic heating by surface (viscoelastic) waves [24, 25].
 The mechanism of anomalous losses in experiments with solid cooled trap surfaces is
 unclear; the key observation is that a trap surface should be as smooth, clean, and
 cool as possible in order to reduce the neutron losses.
 These conditions were satisfied in neutron lifetime measurements [9] performed with a
 low-temperature solid fluoropolymer (a mixture of complex fluoroformaldehydes with a
 molecular weight of ~4500 and a lower melting temperature (~-90 C) than that of Fomblin),
 which was synthesized at the Perm branch of RSC Applied Chemistry and cooled to -160°C.
 The loss coefficient in these experiments was as low as $2\times 10^{-6}$, and the
 difference between the measured neutron storage time and the neutron lifetime obtained
 by extrapolation to the zero frequency of collisions with trap walls was ~ 1\%.
 Further enhancing the accuracy of this method is possible only if the neutron losses in
 traps are reduced to even lower levels (specifically, by reducing the temperature of
 the trap surface).
 Cooling to temperatures significantly lower than the polymer solidification point is
 inefficient: owing to a large temperature expansion coefficient, the deposited solid
 polymer layer cracks and peels off.
 In order to avoid this, a fluoropolymer with a record low melting temperature
 (below -125 C) was proposed in [26].
 The synthesis of this polymer was detailed in [27].
 A coating material with a lower melting point provides the opportunity to store
 ultracold neutrons at a lower temperature and, consequently, with reduced losses at
 thermal vibrations of atoms in the wall.
 The task was to reduce the UCN losses in traps associated with heating by thermal
 vibrations further (i.e., drive them below the levels demonstrated in [9]).
 The geometry of a vertical cylinder with an absorber at the top for varying the
 spectrum of stored neutrons appears to be the best in terms of maximizing the number
 of stored neutrons and enhancing the statistics [28].
 It was found by modelling that this geometry is also the best in terms of simplicity of
 interpreting measurement results [29].
  In the leadup to new measurements of the neutron lifetime in a material trap with the
 above low-temperature fluoropolymer, an experimental setup was constructed and various
 aspects of methodology of future studies (polymer deposition procedure, coating
 stability, reproducibility of the deposition results, variation in the spectrum of
 stored neutrons induced by a special absorber, and coating permeability) were examined.
 Quantitative models had to be constructed for all observations.
 The experiments were conducted at the PF2/UCN platform at the Institut Laue-Langevin
 [30].

\section{Experimental setup}

 The diagram of the setup is presented in Fig. 2.
 The neutron storage tank (6) is a copper cylinder with an inner diameter of 23 cm and a
height of 100 cm (41.5 L in volume).
 Its inner surface is coated with titanium.
Titanium has a small negative potential for neutrons $E_{b}$=-49.7 neV and is a
poor UCN reflector.
 Therefore, the time of neutron storage in a titanium chamber should be no longer than a
 fraction of a second.
 The high absorbing ability of titanium used as a substrate for polymer deposition
 brings the defects of the polymer coating into focus: neutrons penetrating through
 micropores (or penetrating the polymer coating due to its insufficient thickness) enter
 the titanium layer and leave the storage volume.
 The same is true for neutrons with energies exceeding the boundary energy of the
 polymer.
 The cylinder and its base (bottom face of the storage chamber) are cooled by liquid or
 gaseous nitrogen carried by tubes at the bottom and 16 concentric tubes (4) at the
 outer surface of the cylinder.
 A conical copper valve in unit (7) closes and opens the inlet aperture with a diameter
 of 7 cm at the bottom of the storage tank.
 The valve is actuated pneumatically.
 Sputtering device (5) inside the storage chamber is used to apply polymer coating to
 the walls.
 Helium is used as a carrier in the process of sputtering.
 The sputtering device may be shifted in the vertical direction by rod (3) and linear
 drive (1).
 A movable photo camera is used to monitor and image the sputtering process and the
 coating state throughout the entire procedure.  The energy of stored neutrons may be
 altered by introducing an absorber ($^{10}B$) into the storage volume from the top.
 This absorber (5) has the shape of a horizontal disk moved along the vertical axis.
 The neutron valve switch box contains a short section of a neutron guide that connects
 the storage chamber either to the lead in neutron guide supplying neutrons (right
 position in Fig. 2) or to the leadout neutron guide to neutron detector (9) (left
 position in Fig. 2).
 Five small-sized neutron detectors (~l cm in diameter) are located on the side face of
 the chamber at distances of 4, 24, 44, 64, and 84 cm from the bottom.
 The working layer of these detectors is a thin $^{10}B$ layer; charged particles are
 detected by silicon PIPS (passivated implanted planar silicon) detectors (Canberra).
 Side detectors monitor both the processes of filling and emptying the neutron storage
 chamber and the coating quality (by detecting neutrons passing through the film
 deposited onto the surface).
 The storage chamber is cooled by a 200 L dewar with liquid nitrogen.
 Computer-controlled valves provide the opportunity to regulate the flow of liquid or
 gaseous nitrogen in the cooling tubes.
 The temperature of the cylinder and the base is monitored at several points by PT 100
  sensors (IST, Innovative Sensor Technologies).
\section{Coating material and procedure of UCN storage measurements}

The studied fluoropolymer is fluoroformaldehyde
CF$_{3}$(CF$_{2})_{3}$-O-CF$_{2}$-O-(CF$_{2})_{3}$CF$_{3}$  with a relatively low
molecular weight (M = 520) produced by ABCR GmbH (Germany).
 Its synthesis procedure was developed in [27].
 The properties of this polymer are as follows: melting temperature T$_{m}$ =-159.3 C
 and room temperature density in liquid state $\rho$=1.73 g cm$^{-3}$; at temperatures
 close to the nitrogen one, the density increases by ~10\% (these data were provided by
 the manufacturer and, in part, obtained in a laboratory at the Johannes Gutenberg
 University Mainz).
 This low-temperature density corresponds to the potential of ~105 neV for neutrons.
 In contrast to the wall materials used in room-temperature UCN storage with coatings
 applied in an open chamber under atmospheric pressure, the deposition of a
 low-temperature polymer with a high vapor pressure at room temperature is carried out
 either by evaporation [9] or by sputtering (as in the present study).

 The procedure of coating and UCN storage measurements was as follows.
 Liquid polymer was introduced into a glass vessel, air was evacuated, and the polymer
 was heated to boiling; in the process, pure helium was supplied into the vessel and
 evacuated via a flexible small-diameter hose to the UCN storage chamber.
 The polymer condensed in the tube and was sputtered in the form of small droplets onto
 the chamber surface.
 The exit section of the hose with a nozzle was moved vertically to produce a uniform
 coating.
 Polymer sputtering was performed at a near-nitrogen temperature of the chamber walls.
 The polymer was thus deposited in the form of snow or loose ice with a considerable
 number of defects onto the surface (Fig. 3).  When the temperature was increased later
 in the process, the polymer melted and coated the surface uniformly.
 Subsequent rapid cooling resulted in solidification and the formation of a smooth
 coating.
 The complete measurement cycle (filling the chamber with neutrons, storing them for a
 certain amount of time, and emptying the chamber to the detector) was repeated multiple
 times within this temperature cycle.
 This is illustrated in Fig. 4, where astronomical time is on the horizontal axis, the
 number of neutrons remaining in the chamber after a certain storage time is plotted on
 the left vertical axis, and temperature is plotted on the right vertical axis.
 The numbers next to points indicate the storage time: 5, 50, 100, 300, and 500 s.
 The number of neutrons decreases slightly with time.
 The reason for this is unclear; it may be caused by the precipitation of water vapor
 from the atmosphere due to the imperfect vacuum.
 It can be seen from Fig. 4 that the number of stored neutrons in experiments with an
 as-deposited coating at nitrogen temperature is very small, which is indicative of a
 low coating quality.
 This quality increases as the temperature rises, and the number of stored ultracold
 neutrons also increases and reaches its maximum at -145 C.
 At this moment, an abrupt temperature drop was induced, the polymer coating solidified,
 and the principal series of measurements was performed.
 Temperature variations within 10 C did not affect the UCN accumulation and storage.
 The number of neutrons that accumulated in typical polymer deposition cycles was as
 high as ~$5\times 10^{4}$, which
corresponds to a chamber average density of ~1 cm$^{-3}$.

\section{Modelling}

 Since the experimental data are supplemented with a quantitative interpretation, we
 first detail the formalism for this interpretation.
 In accordance with the adopted formalism [3], velocity-dependent probability
 <$\mu(v)$> (averaged over an isotropic angular distribution) of losing a neutron in a
 collision with a thick wall impenetrable for UCN is written as
\begin{equation}
<\mu(v)>=2\,\eta\frac{\kappa(y)}{y^{2}},\quad
\kappa(y)=\arcsin{y}-y\sqrt{1-y^{2}},\quad y=v/v_{b}=\sqrt{E/V},
\end{equation}
where $v$ and $E$ are the velocity and the energy of a neutron, $v_{b}$ is the boundary
wall velocity, and $\eta$ is the UCN loss coefficient defined by complex potential $U$
that depends on the properties of the wall material:
\begin{equation}
\eta=-\mbox{ Im }U/\mbox{ Re }U, \quad U=(\hbar^{2}/2m)4\pi\sum_{i}N_{i}b_{i},
\quad V=\mbox { Re }U,\quad\mbox{ Im }b=-\sigma/2\lambda.
\end{equation}
Here $m$ is the neutron mass, $N_{i}$ is the atomic density of component $i$ of the wall material, $b_{i}$ is the corresponding coherent length of scattering by bound nuclei, and $\sigma$ is the cross section of inelastic processes (capture and heating) for neutrons with wavelength $\lambda$.

 We follow the formalism [31] in modelling the UCN behavior in storage chamber with the
 gravity field taken into account.
 The neutron loss rate (s$^{-1}$)  in the gravity field at an equilibrium UCN
 distribution in the available phase space is given by
\begin{equation}
d\int \varrho({\bf r,v},t)d^{3}{\bf r}\;d^{3}{\bf v}=
-\int\mu({\bf v})({\bf v}{\bf n_{S}})\varrho({\bf r},{\bf v},t)\;
d^{3}{\bf v}\;dS\;dt.
\end{equation}
Here, $\varrho({\bf r},{\bf v},t)$  is the UCN density in the phase space and
${\bf n_{S}}$ is the normal to the surface at the point of collision.
 The neutron density depends on height $z$ measured from the lowest point of the
 storage chamber:
\begin{equation}
\varrho(z,v)=c\; \delta(v_{0}^{2}-2gz-v^{2}),
\end{equation}
where $c$ is the normalization constant, $v_{0}$ is the velocity of a neutron at the
bottom of the chamber, $v$ is the velocity of a neutron at height $z$,
$\varrho(0,v)=c\;\delta(v_{0}^{2}-v^{2})$ is the neutron density at the bottom of the
chamber, and
$g=9,80665$\,m s$^{-2}$ is the acceleration due to gravity.

 In vertical-cylinder geometry, the integral at the left-hand side of Eq. (3)
 (effective volume) is transformed into
\begin{equation}
c(\pi r)^{2}\int_{0}^{v_{0}^{2}/2g}(v_{0}^{2}-2gz)^{1/2}dz=c\frac{2(\pi
r)^{2}v_{0}^{3}}{3g},
\end{equation}
where $r$ is the cylinder radius.

The right-hand side is determined by the losses in the cylinder wall:
\begin{equation}
c\pi^{2}r\int_{0}^{v_{0}^{2}/2g}(v_{0}^{2}-2gz)<\mu (v)>dz,
\end{equation}
and the losses at the bottom of the chamber: $c(\pi r)^{2}<\mu (v_{0})>v_{0}^{2}/2$.

 Thus, the rates of neutron loss in collisions with the side wall and the bottom face of
the chamber are given by
\begin{equation}
\frac{3g}{2rv_{0}^{3}}\int_{0}^{v_{0}^{2}/2g}(v_{0}^{2}-2gz)<\mu (v)>dz \quad
{\mbox and}\quad  \frac{3g<\mu (v_{0}>}{4v_{0}}.
\end{equation}
 The UCN losses or the UCN transmission probability may be calculated in a similar way
 for any part of the chamber surface; integration in Eq. (7) is then performed within
 the needed coordinate interval.
 For example, the neutron leak rate through a gap possibly existing at the bottom of the
 chamber (specifically, in the neutron inlet-outlet valve) is written as
 $(3g<\mu (v_{0})>s)/(4v_{0}S)$, where $s$ and $S$ are the areas of the gap and the
 bottom face, respectively.
 In the case of spectral UCN distributions $\rho(v_{0})$, Eq. (7) is integrated over
 these distributions.
 The experimental UCN storage curves in the chamber with a fluoropolymer coating
 revealed UCN transmission through the coating (manifested in the count rate of side
 counters).
 The experimentally observed UCN storage in the titanium chamber without a special
 fluoropolymer coating may be attributed to the presence of an oxide film and, possibly, a thin fluoropolymer film.
 In order to interpret the UCN storage in the chamber with a multilayer coating, one
 should calculate the UCN transmission through such multiplayer films.
 This was done using the formalism [32] of recurrence formulas containing analytical
 solutions for the amplitudes of transmitted and reflected waves for an arbitrary
 sequence of complex potentials.

 The amplitudes of transmission and reflection of neutron waves from the $n_{th}$
 boundary in the potential sequence take the form
\begin{equation}
T_{n}=\frac{t_{n}^{+}e^{i\phi_{n-1}}T_{n-1}}
{1-r_{n}^{-}R_{n-1}e^{2i\phi_{n-1}}}
\end{equation}
and
\begin{equation}
R_{n}=\frac{r_{n}^{+}+(t_{n}^{+}t_{n}^{-}-r_{n}^{+}r_{n}^{-})
e^{2i\phi_{n-1}}R_{n-1}}{1-r_{n}^{-}R_{n-1}e^{2i\phi_{n-1}}}
\end{equation}
respectively, where
\begin{equation}
t_{n}^{+}=\frac{2k_{n}}{k_{n}+k_{n-1}}; \quad
t_{n}^{-}=\frac{2k_{n-1}}{k_{n}+k_{n-1}}; \quad
r_{n}^{+}=-r_{n}^{-}=\frac{k_{n}-k_{n-1}}{k_{n}+k_{n-1}}; \quad
k_{i}=\sqrt{k^{2}-u_{i}}.
\end{equation}
Here, $k$ is the wave vector of a neutron in vacuum,  $u_{i}=2mU_{i}/\hbar^{2}$,
$\phi_{i}=k_{i}l_{i}$,
and $l_{i}$ is the thickness of layer $i$ in the potential sequence.
 The probabilities of transmission and reflection from a sequence of $n$ potentials are
 $w_{t}=|T_{n}|^{2}$ and  $w_{r}=|R_{n}|^{2}$ respectively.
 Angle-averaged coefficients of transmission and reflection $<\mu (v)>$ for such
 layered coatings were calculated for isotropic angular distributions of incident
 neutrons.

\section{Experiments with the Boron absorber used to shape the initial UCN spectrum}

 The initial UCN spectrum was altered using boron ($^{10}B$ with an enrichment of 94\%)
 absorber (5) (see Fig. 2).
 When the chamber is filled, neutrons reaching the absorber surface in the gravity field
 are absorbed with a high probability; after a certain interval, only the neutrons with
 the upper spectrum boundary set by the height of the absorber position $E_{b}$\,
 (neV)=1,025 h$_{abs}$\,( cm) should remain in the chamber.
 Probability R of neutron reflection from the surface with complex potential $U$ is
 given by
\begin{equation}
R=\Biggl|\frac{k_{0,\perp}-k_{\perp}}{k_{0,\perp}+k_{\perp}}\Biggr|^{2}.
\end{equation}
Here, $k_{0,\perp}$ is the component of the neutron momentum in vacuum normal to the
surface, $k_{\perp}=(k_{0,\perp}^{2}-2mU/\hbar^{2})^{1/2}$ is the corresponding
component in the medium, $U=\frac{\hbar^{2}}{2m}\sum_{i} 4\pi N_{i}b_{i}$, N$_{i}$ is
the atomic density, and b$_{i}$ is the complex coherent scattering length of a neutron
in the medium.
 The calculated coefficients of UCN reflection (under isotropic incidence) from the
 surface of $^{10}$B and other absorbers (titanium, polyethylene, and natural gadolinium)
 are presented in Fig. 5.
 Complex potential (2) used in the calculations of reflection from the surface of
 enriched boron is Re U= 6.624 neV, ImU =-31.38 neV.
 The complex potential of gadolinium is Re U = 74.9 neV, Im U =-108.8 neV.
 The absorption coefficients of pure $^{10}$B and $^{10}$B$_{0,94}\,\,^{11}$B$_{0,06}$
 (94\% enrichment) are almost equal.

 Figure 6 presents the experimental data on UCN storage in the chamber with a
 fluoropolymer coating at different positions of the boron absorber.
 The curves approximating these data in the model with two exponential decay constants
 are also shown in Fig. 6.
 The fast decay component corresponds to the absorption of neutrons with energy
 E > mgh$_{abs}$ exceeding the one needed to reach the absorber in the gravity field,
 while the slow component represents the process of storage of neutrons with lower
 energies. The curves have a qualitatively natural shape: when the absorber moves
 downward (i.e., the energy of stored neutrons decreases), the faster component of the
 neutron spectrum is absorbed faster, while the storage time of the slow (unabsorbed by
 boron) spectrum component increases, since the UCN losses are lower for neutrons with
 lower energies.
 The numbers of neutrons unabsorbed by boron determined at different heights of the
 absorber position and extrapolated to the initial moment are of interest.
 These numbers characterize the shape of the initial neutron density spectrum
 $\varrho(E)$ in the storage chamber: $N=\int\varrho(E)dE$ in the interval from zero to
 the energy set by the absorber position.
 Figure 7 presents these data together with the curves calculated under different
 assumptions regarding the shape of this initial neutron density spectrum:
 $n(E)\sim E^{d}$ at $d$=1/2 (Maxwellian density spectrum), d=1, and d=3/2.
 It can be seen that the experimental data correspond to a spectrum that is depleted at
 low neutron energies relative to the Maxwellian spectrum $n(E)\sim E^{1/2}$.
 The best fit is obtained for $n(E)=2.2 E^{1.2}/E_{max}^{2.2}$ (the spectrum is
 normalized to unity).
 All subsequent calculations of UCN storage were carried out for this neutron spectrum.

\section{UCN storage in Titanium chamber without fluoropolymer coating}

 The low coefficient of neutron reflection from a surface with a negative potential of
interaction with neutrons (U$_{Ti}$=-49.7 neV) makes the UCN storage in the
titanium  chamber impossible (Fig. 5).
 However, the presence of impurities, an oxide film with a positive potential
(U$_{TiO_{2}}$=68.9 neV), or a thin film of an incompletely removed
fluoropolymer may induce noticeable neutron reflection.
 The variation in the number of neutrons stored in the titanium chamber with time was
 determined by monitoring the count rate of five counters on the side face of the
 chamber immediately after the completion of 40 s filling chamber.

 Figure 8 shows the count rate of the side neutron counter located 4 cm from the bottom
 face.
 It can be seen that the storage of neutrons in the titanium chamber is characterized
 by a decay time of ~3 s.
 If one tries to explain the measured storage time by the presence of an oxide film only
(in an assumption of thick titanium layer), a TiO$_{2}$ film thickness of
~500\,\AA\, will be needed to reproduce the experimental data in the modelling
of temporal evolution of the number of stored neutrons (Fig. 8).
 This high thickness of the  TiO$_{2}$ film thickness seems unphysical.
 Therefore more complicated potential sequence was tried for describing the UCN storage
 in Ti chamber : TiO$_{2}$, Ti, and Cu - reflective wall behind the Ti layer.


\section{UCN storage in the chamber with a fluoropolymer coating}

 The shapes of potentials of the chamber walls used in the UCN reflection calculations
are presented in Fig. 9.
 A sequence of four potentials was used to calculate the process of UCN storage in the
fluoropolymer-coated chamber: (1) fluoropolymer, U=106 neV; (2) titanium
dioxide TiO$_{2}$, Re U=68.9 neV, Im U=-1.4$\times 10^{-2}$\,neV;  (3)
titanium, Re U=-49.7 neV, Im U=-2.5$\times 10^{-2}$\,neV, and copper: Re U=170
neV, Im U=-2.31$\times 10^{-2}$\,neV (see the left part of Fig. 9).
 The right part of Fig. 9 demonstrates the sequence of potentials in calculations of
the count rate of boron counters on the side face of the storage chamber: (1)
fluoropolymer; (2) enriched boron layer.
 Neutrons are incident from the right; the thicknesses are not to scale.

 The interpretation of data on UCN storage in the fluoropolymer-coated titanium
chamber is ambiguous, since different combinations of thicknesses of  TiO$_{2}$, Ti,
and fluoropolymer films satisfy the experimental UCN storage curve.
 The count rate of side counters was used to obtain definite values of thickness of all
films.
 At a TiO$_{2}$ film thickness of ~500 A (determined by modelling the UCN storage in the
titanium chamber), the modelling of neutron storage in the fluoropolymer-coated
chamber requires a polymer coating with a thickness of just ~365 A for the
results to agree with the experimental storage curve (see Fig. 11).
 However, the corresponding calculated count rate of side counters with an active
$^{10}B$ layer, which should also have a fluoropolymer coating on the surface,
is more than an order of magnitude higher than the measured one (Fig. 10).
 Various hypothetical combinations of thicknesses of the TiO$_{2}$ film and the residual
fluoropolymer (in UCN storage in the titanium chamber) or the fluoropolymer (in
storage in the chamber with a special coating) film were tested to bring these
discrepant results into agreement.
 It should be noted that the reproducibility of UCN storage curves in the sputter-coated
chamber was low: the spread of storage times obtained in different cycles of
polymer deposition exceeded 50\%.
 As was mentioned above, UCN penetration through the polymer coating was observed.
 Figure 10 shows the measured count rate of the side boron counter as a function of the
storage time in experiments performed in the fluoropolymer-coated chamber.
 The approximation with a single-exponential model (solid curve) provides the following
time dependence of the count rate: $N(t)=(0.024\pm 0.11)+(0.80\pm 0.23)\times
exp(-t/(157\pm 98))$.
 This count rate fits the model if the thickness of the polymer coating is 650\,\AA.
 Figure 11 (dots) shows an example of an experimental UCN storage curve (i.e., variation
in the number of neutrons remaining in the chamber with storage time).
 If the observed transmission of stored neutrons through the polymer coating is
neglected and the obtained result is interpreted in terms of Eqs. (1) and (2)
under different assumptions regarding the value of loss coefficient $\eta$ for
the fluoropolymer with E$_{b}=106$\,neV, the best fit is obtained at $\eta\sim
2.7\cdot 10^{-4}$, which is two orders of magnitude higher than the expected
value of $\eta\sim 2^\times 10^{-6}$.
 The results of modeling of the UCN storage with an assumed fluoropolymer thickness of
650\,\AA and two different thicknesses of the TiO$_{2}$ film (100 and 120\,\AA)
are also presented in Fig. 11.
 The best fit is obtained under the assumption that the $TiO_{2}$ thickness falls within
the 110-120\,\AA range.
 All experimental data are brought into agreement only under the assumption that the Ti
coating thickness is 4000-5000\,\AA, the thickness of the TiO$_{2}$ oxide film
is 110 A, and the thickness of the fluoropolymer coating is 650\,\AA.

\section{Conclusions}

The experiments on UCN storage in the chamber coated with a low-molecular-weight fluoropolymer in solid state at a low temperature revealed the following.
\begin{itemize}
\item
 The procedure of sputter coating deposition did not yield reproducible results.
\item
 The penetration of neutrons through the polymer coating was observed.
 This may be caused by the insufficient coating thickness, which is probably attributable to the polymer flow due to its low viscosity in the liquid state after melting before the subsequent freezing.
\item
 The algorithms for calculating the process of UCN storage in the chamber with a wall characterized by a layered complex potential were developed.
 The behavior of neutrons in different experimental conditions was modeled.
 The results of this modeling provided the opportunity to bring the data of various measurements (UCN storage in the titanium chamber, UCN storage in the fluoropolymer-coated chamber, spectrum alterations induced by the boron absorber, and count rates of boron UCN counters on the side face of the storage chamber) into agreement.
\item
 The constructed setup for studies into UCN storage in low-temperature chambers may well
 be used (after some upgrade) in further research into UCN storage in chambers coated with fluoropolymers at low temperatures.
\end{itemize}

\section{Acknowledgments}
We thank the reactor personnel at the Institut Laue-Langevin for maintaining smooth reactor operation and their help.


REFERENCES

l. D. D\"ubbers and M. C. Schmidt, Rev. Mod. Phys. 83, 1111 (2011).

2. F. E. Wietfeldt and G. L. Greene, Rev. Mod. Phys. 83, 1173 (2011).

3. Ya. B. Zel'dovich, ZhETF 36, 1952 (1959);  Sov. Phys. JETP 9, 1389 (1952);

F. L. Shapiro, in Proceedings of the International Conference on Nuclear Structure with Neutrons, Budapest, 1972,
Ed. by J. Ero and J. Szucs (Plenum, New York, 1972), p. 259;

A. Steyerl, Springer Tracts Mod. Phys. 80, 57 (1977);

R. Golub and J. M. Pendlebury, Rep. Prog. Phys. 42, 439 (1979);

V. K. Ignatovich, The Physics of Ultracold Neutrons (Nauka, Moscow, 1986; Clarendon, Oxford, 1990);

R. Golub, D. J. Richardson, and S. Lamoreaux, Ultracold Neutrons (Adam Hilger, Bristol, 1991);

 J. M. Pendlebury, Ann. Rev. Nucl. Part. Sci. 43, 687 (1993).

4. J. Byrne et al., Europhys. Lett. 33, 187 (1996).

5. J. S. Nico et al., Phys. Rev. C 71, 05550 (2005).

6. A. T. Yue et al., Phys. Rev. Lett. 111, 222501 (2013).

7. A. Pichlmaier, et al., Nucl. Instrum. Methods Phys. Res, Sect. A440, 517 (2000).

8. A. Pichlmaier et al., Phys. Lett. B 693, 221 (2010).

9. A. Serebrov et al., Phys. Lett. B 605, 7 (2005); A. Serebrov et al., Phys. Rev. C 78, 035505 (2008).

10. S. Arzumanov et al., Phys. Lett. B 483, 15 (2000);
S.Arzumanov et al., Nucl. Instrum. Methods Phys. Res, Sect. A440, 511 (2000).

11. S. Arzumanov et al., JETP Lett. 95, 224 (2012).

12. S. Arzumanov et al., Phys. Lett. B 745, 79 (2015).

13. V. F. Ezhov, in Proceedings of the International Conference on Ultracold and Cold Neutrons Physics and Sources, St-Petersburg, 2009;
V. F. Ezhov et al., arXiv:1412.7434[nucl-ex];
V.F. Ezhov et al., Pis'ma v ZETF 107, 707 (2018).

14. C.L. Morris et al., LA-UR-16-27352 (Los Alamos Natl. Lab., 2016); arXive:1610.04560[nucl-ex].

15. R.W. Pattie, Jr. et al., arXive:1701.01817 [nucl-ex].

R.W. Pattie, Jr. et al.,  Science 360, 627 (2018).

16. A. P. Serebrov, E. A. Kolomenskiy, A. K. Fomin, et al., JETP Lett. 106, 623
(2017).

17. A. Steyerl, S. S. Malik, A. M. Desai, and C. Kaufman, Phys. Rev. C 81, 055505 (2010).

18. A. Steyerl, J. M. Pendlebury, C. Kaufman, S. S. Malik,
and A. M. Desai, Phys. Rev. C 85, 065503 (2012).

19. Z. Berezhiani and L. Bento, Phys. Rev. Lett. 96, 081801 (2006);

 Z. Berezhiani, Eur. Phys. J. 64, 421 (2009);

Z. Berezhiani, arXiV:1602.08588[astro-ph.];

 Z. Berezhiani, in Proceedings of the International Workshop on Probing Fundamental
 Symmetries and Interactions with UCN, Mainz, Germany, Apr. 11-15, 2016.

20. M. Sarazzin et al., Phys. Lett. B 712, 213 (2012);

M. Sarazzin et al., Phys. Lett. B 758, 14 (2016).

21. J. C. Bates, Phys. Lett. 88A, 427 (1982); J. C. Bates,
Nucl. Instrum. Methods Phys. Res, Sect. A 216, 535
(1983).

22. Yu. N. Pokotilovski, J. Exp. Theor. Phys. 96, 172
(2003).

23. www.solwaysolexis.com.

24. Yu. N. Pokotilovski, Phys. Lett. A255, 173 (1999).

25. S. K. Lamoreaux and R. Golub, Phys. Rev. C 66, 044309 (2002).

26. Yu. N. Pokotilovski, Nucl. Instrum. Methods Phys. Res, Sect. A425, 320 (1999).

27. K. Sung and R. J. Lagow, J. Am. Chem. Soc. 117, 4276 (1995);

K. Sung and R. J. Lagow, Synth. Commun. 26, 375 (1996).

28. Yu. Yu. Kosvintsev, V. I. Morozov, and G. I. Terekhov, JETP Lett. 36, 424
(1982).

29. Yu. N. Pokotilovski, Phys. At. Nucl. 73, 725 (2010).

30. www.ill.fr/YellowBook/PF2/.

31. V. K. Ignatovich and G. I. Terekhov, JINR Comm. R4-9567 (JINR, Dubna, 1976).

32. V. K. Ignatovich, JINR Preprint No. P4-10778 (JINR, Dubna, 1977); JINR Preprint
No. PR4-11135 (JINR, Dubna, 1978).


\begin{figure}
\begin{center}
\resizebox{13cm}{13cm}{\includegraphics[width=\columnwidth]{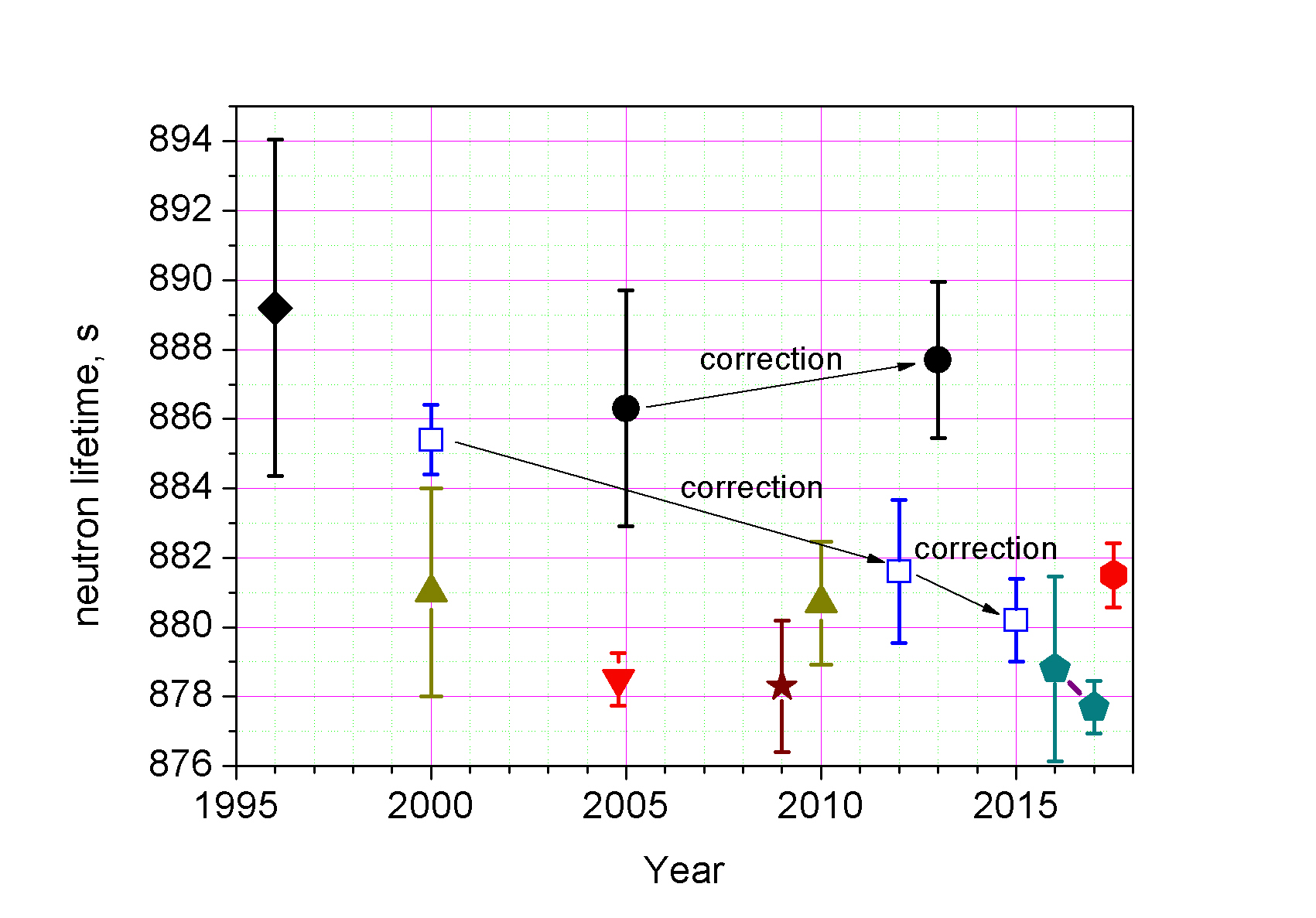}}
\end{center}
\caption{ Fig. 1. Summary of the results of neutron lifetime measurements
performed in the last 20 years: beam experiment[4] (diamond), beam experiments
[5, 6] (filled circles), experiments [7, 8] on UCN storage in the chamber
coated with liquid Fomblin (upward-pointing triangles), experiment [9] on UCN
storage in the chamber coated with frozen PFPOM (downward-pointing triangle),
UCN storage in the magnetic trap [13] (star symbol), UCN storage with the
detection of heated neutrons leaving the chamber [10-12] (squares), UCN storage
in the magnetic trap[14, 15] (pentagon), and experiment [16] on UCN storage in
the chamber coated with frozen Fomblin grease (hexagon).}
\end{figure}

\newpage
\begin{figure}
\begin{center}
\resizebox{13cm}{13cm}{\includegraphics[width=\columnwidth]{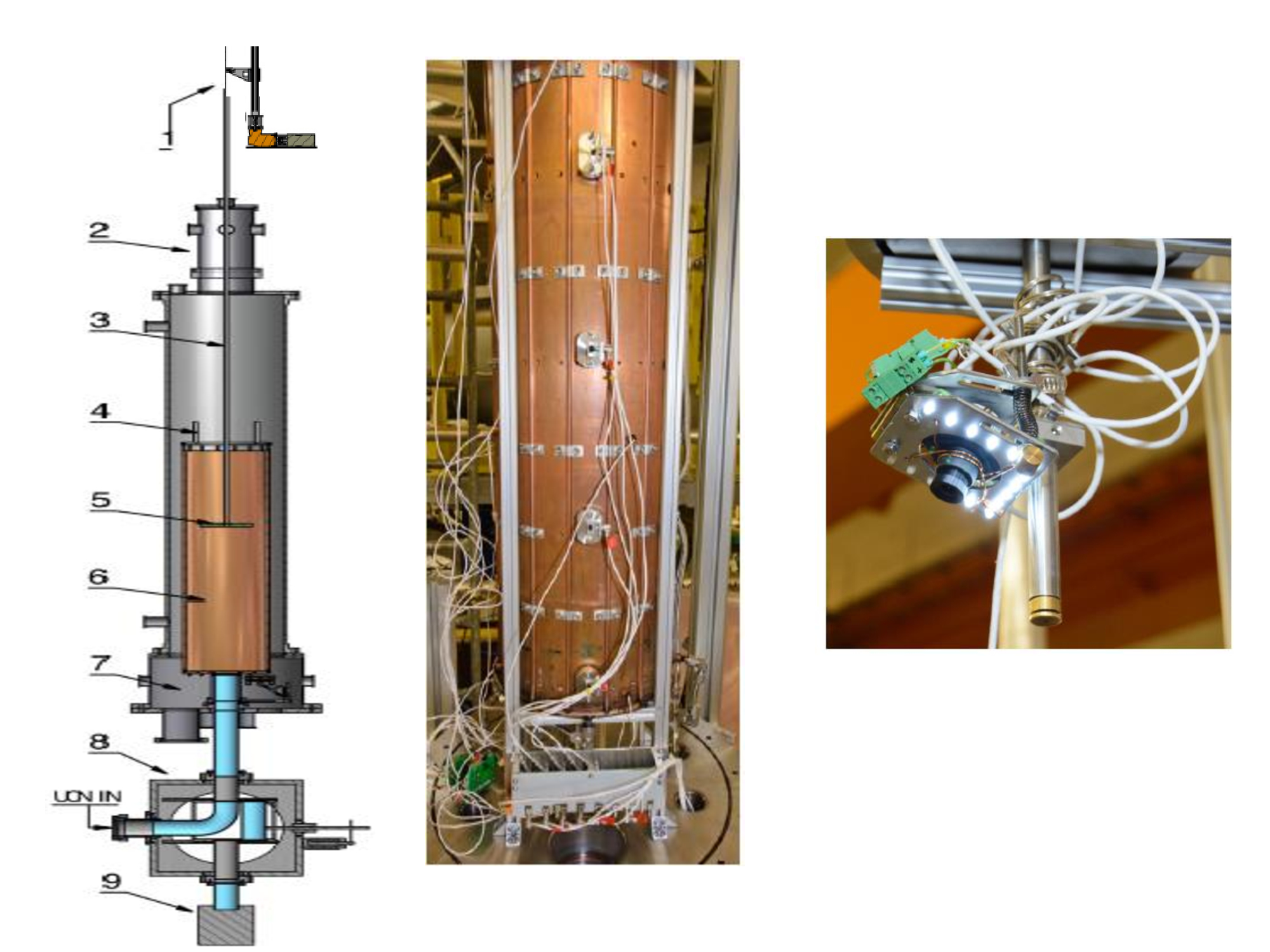}}
\end{center}
\caption{Diagram of the setup. Left: (1) linear drive, (2) vacuum valve and
seal, (3) rod that moves the sputtering device and the boron absorber
vertically, (4) cooling tubes on the side face, (5) polymer sputtering device
and boron absorber, (6) UCN storage volume, (7) neutron valve unit, (8) neutron
valve switch box, and (9) neutron detector.
 Center: Photographic image of the setup with the vacuum jacket removed.
 Vertical cooling tubes and side neutron detectors on the exterior surface of the
 storage chamber, the bottom of the chamber, and the cable layout are visible.
 Right: photo camera.}
\end{figure}

\newpage
\begin{figure}
\begin{center}
\resizebox{13cm}{13cm}{\includegraphics[width=\columnwidth]{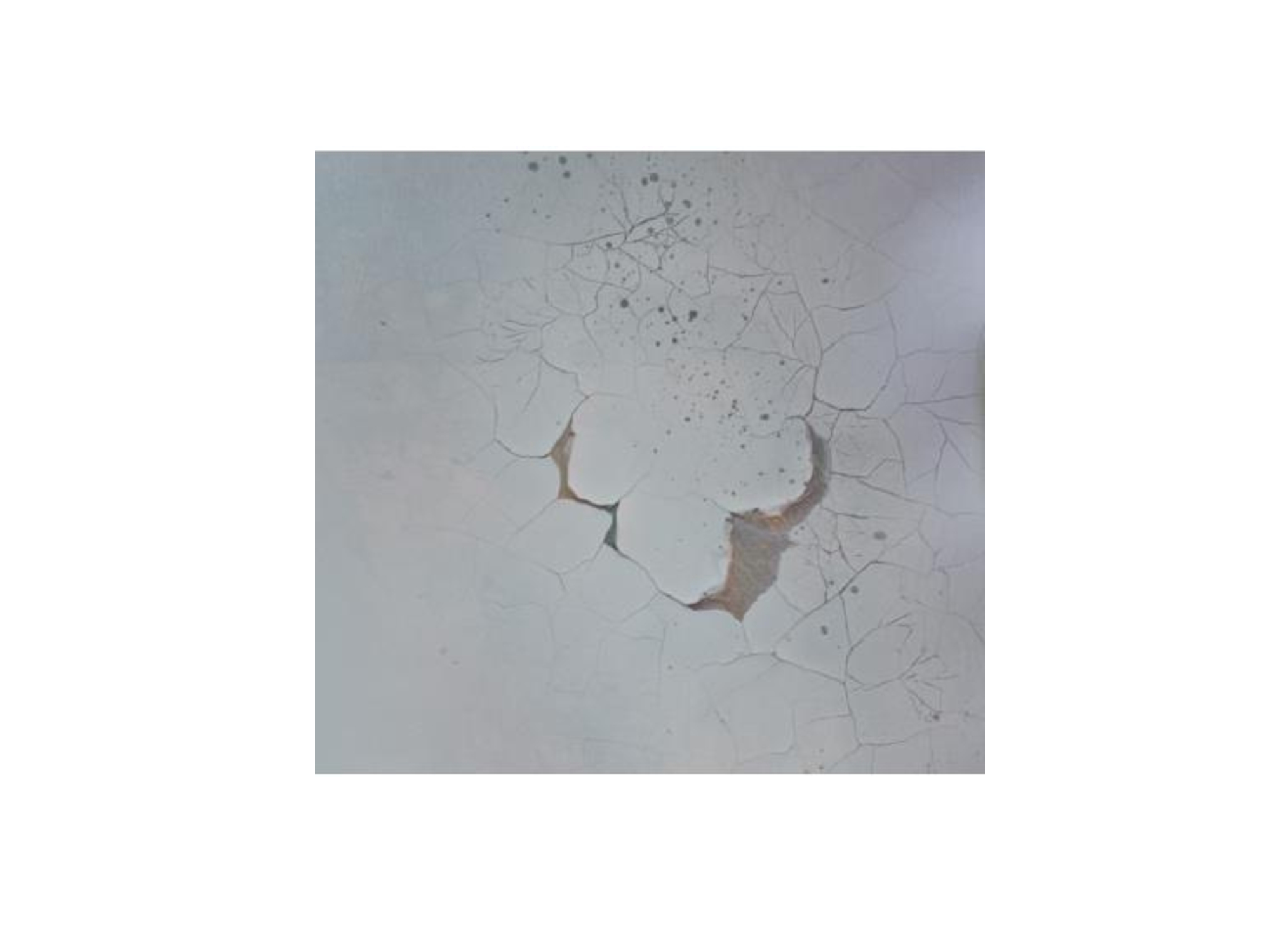}}
\end{center}
\caption{Surface of the fluoropolymer coating immediately after its deposition onto the
surface of the titanium chamber at liquid nitrogen temperature.}
\end{figure}

\newpage
\begin{figure}
\begin{center}
\resizebox{13cm}{13cm}{\includegraphics[width=\columnwidth]{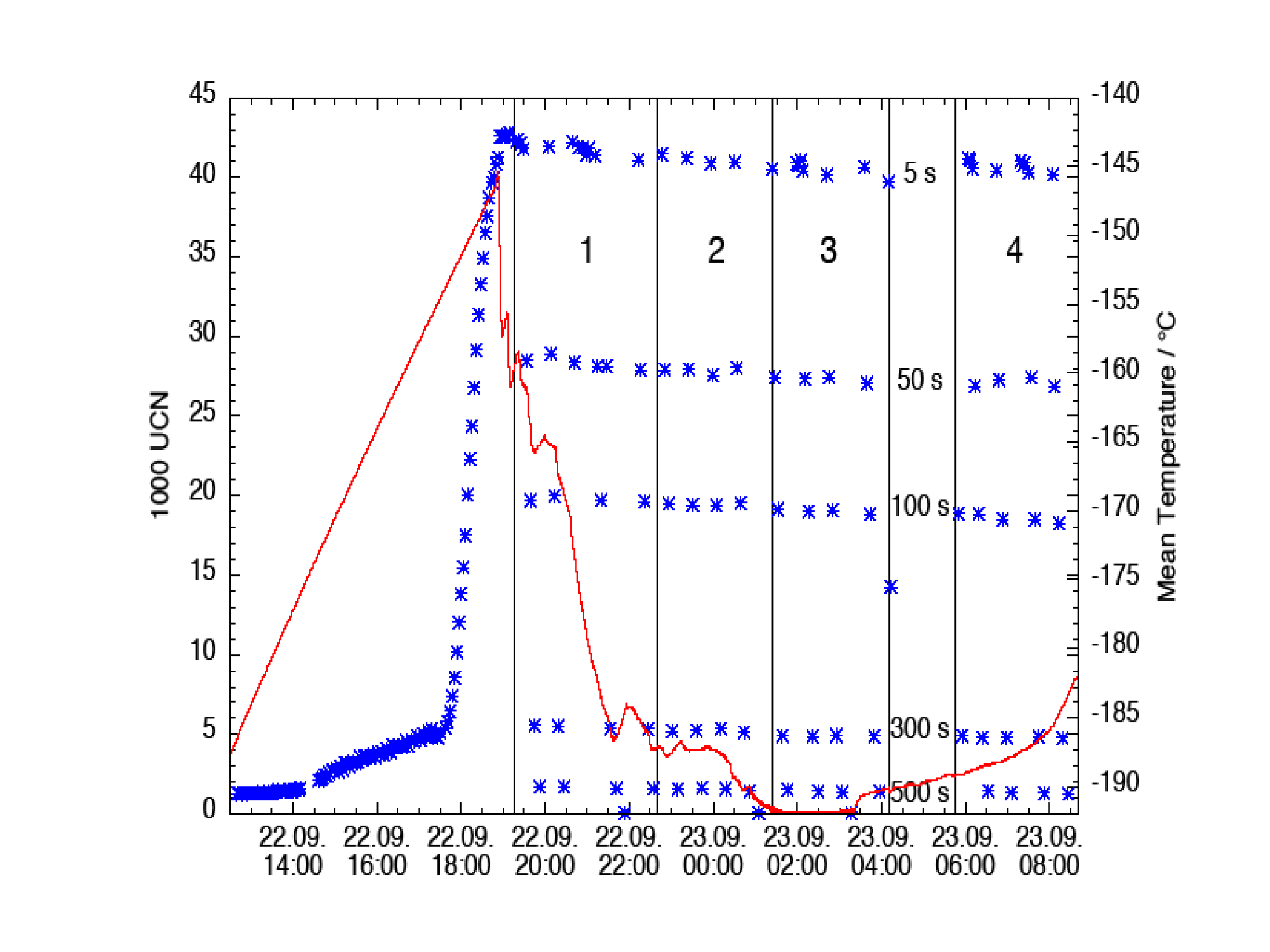}}
\end{center}
\caption{Temporal pattern of UCN storage in the fluoropolymer-coated chamber.
 The curve represents the variation of temperature with time (right scale).
 Dots correspond to the numbers of neutrons remaining after 5, 50, 100, 300, and 500 s
 (left scale).
 The astronomical measurement time is indicated on the horizontal scale.}
\end{figure}

\newpage
\begin{figure}
\begin{center}
\resizebox{13cm}{13cm}{\includegraphics[width=\columnwidth]{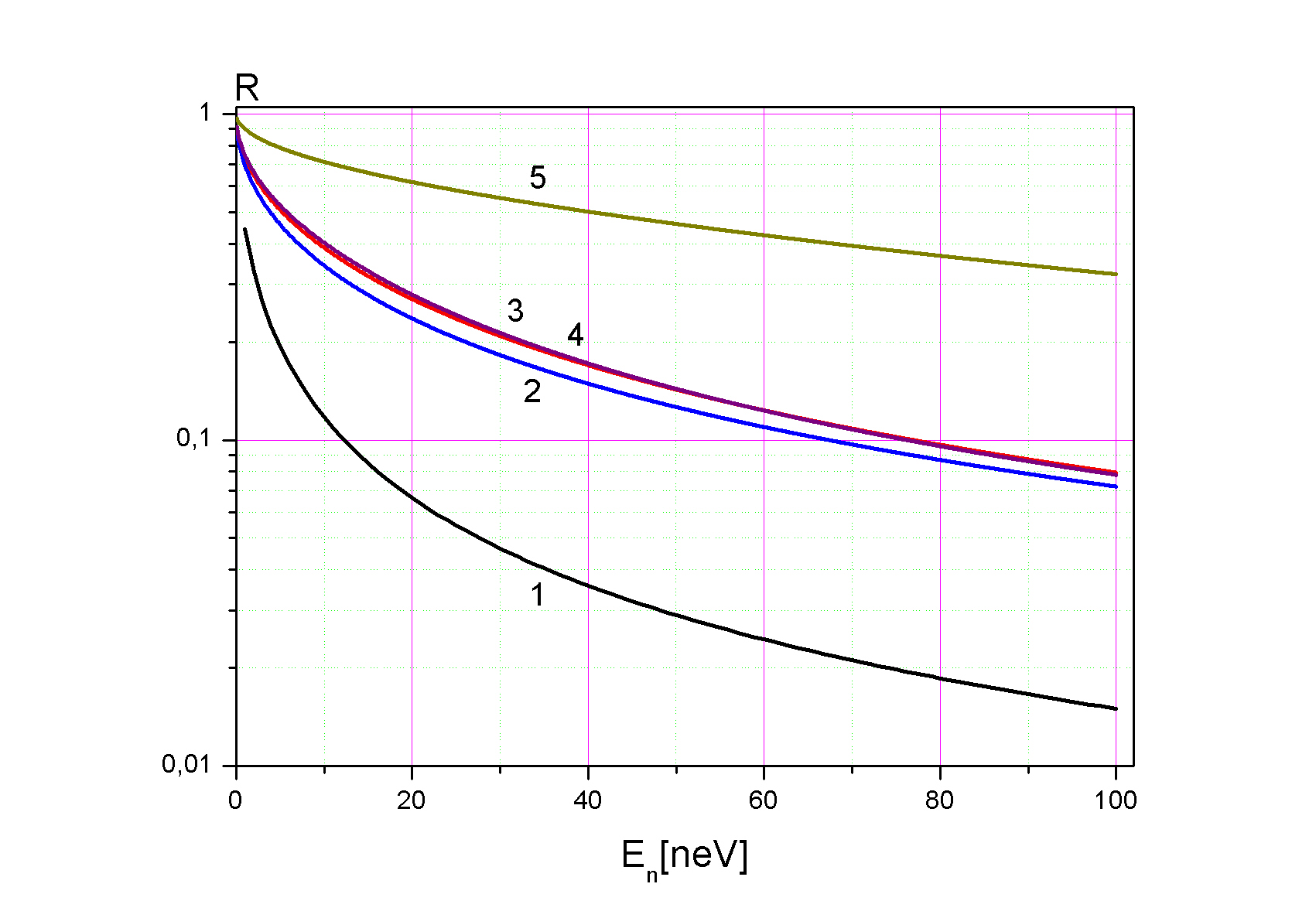}}
\end{center}
\caption{Calculated coefficients of reflection of neutrons incident
isotropically onto the surface of several absorbers: polyethylene (1), titanium
(2), $^{10}B$, and $^{10}B_{0.94}\,^{11}B_{0.06}$ (3, 4), and gadolinium (5).}
\end{figure}

\newpage
\begin{figure}
\begin{center}
\resizebox{13cm}{13cm}{\includegraphics[width=\columnwidth]{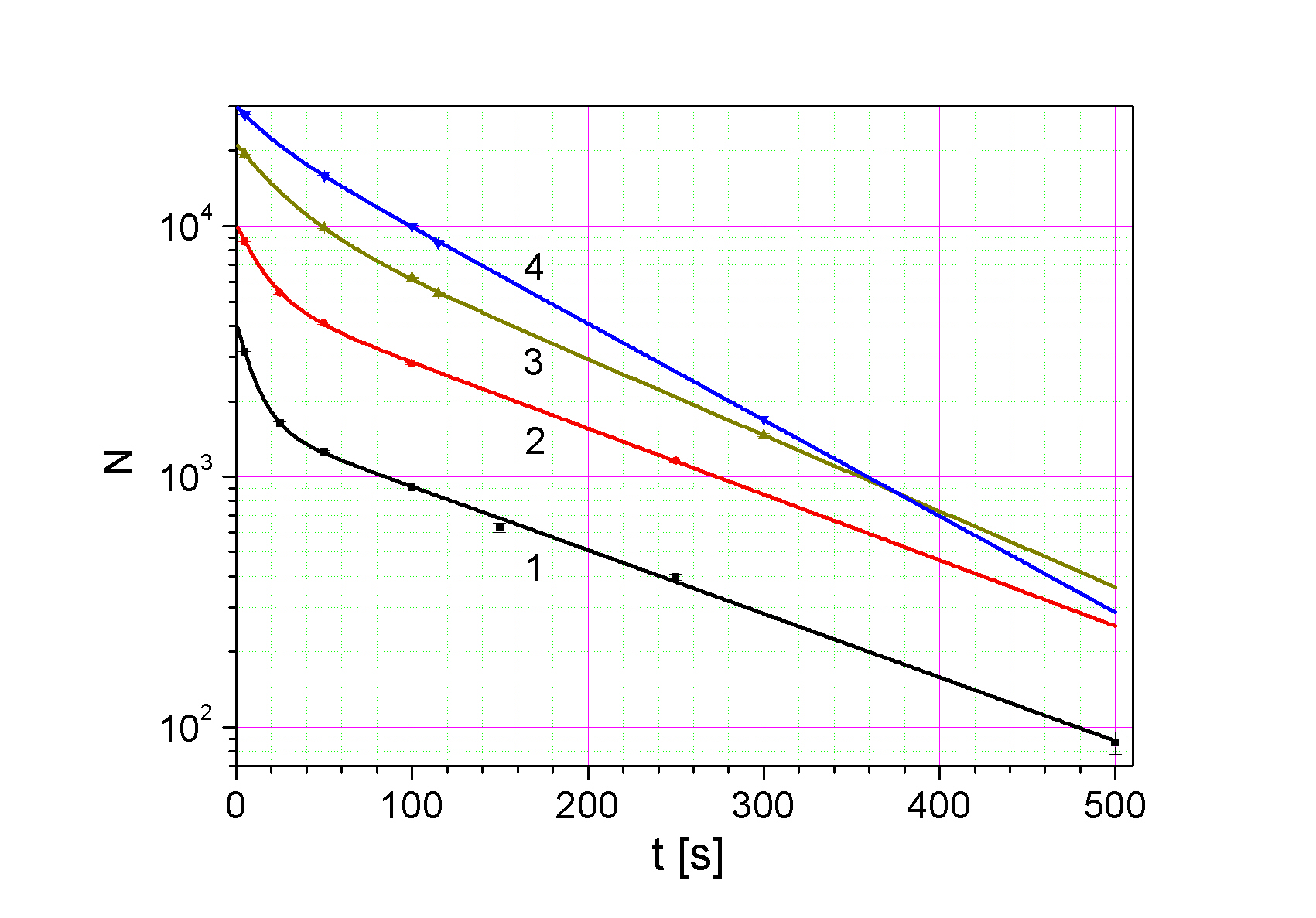}}
\end{center}
\caption{Measured UCN storage curves in the fluoropolymer-coated chamber with a boron
absorber located at a height of 31 (1), 50(2), 75 (3), or 100 cm (4).
 The approximating curves were obtained in a double-exponential model.}
\end{figure}

\newpage
\begin{figure}
\begin{center}
\resizebox{13cm}{13cm}{\includegraphics[width=\columnwidth]{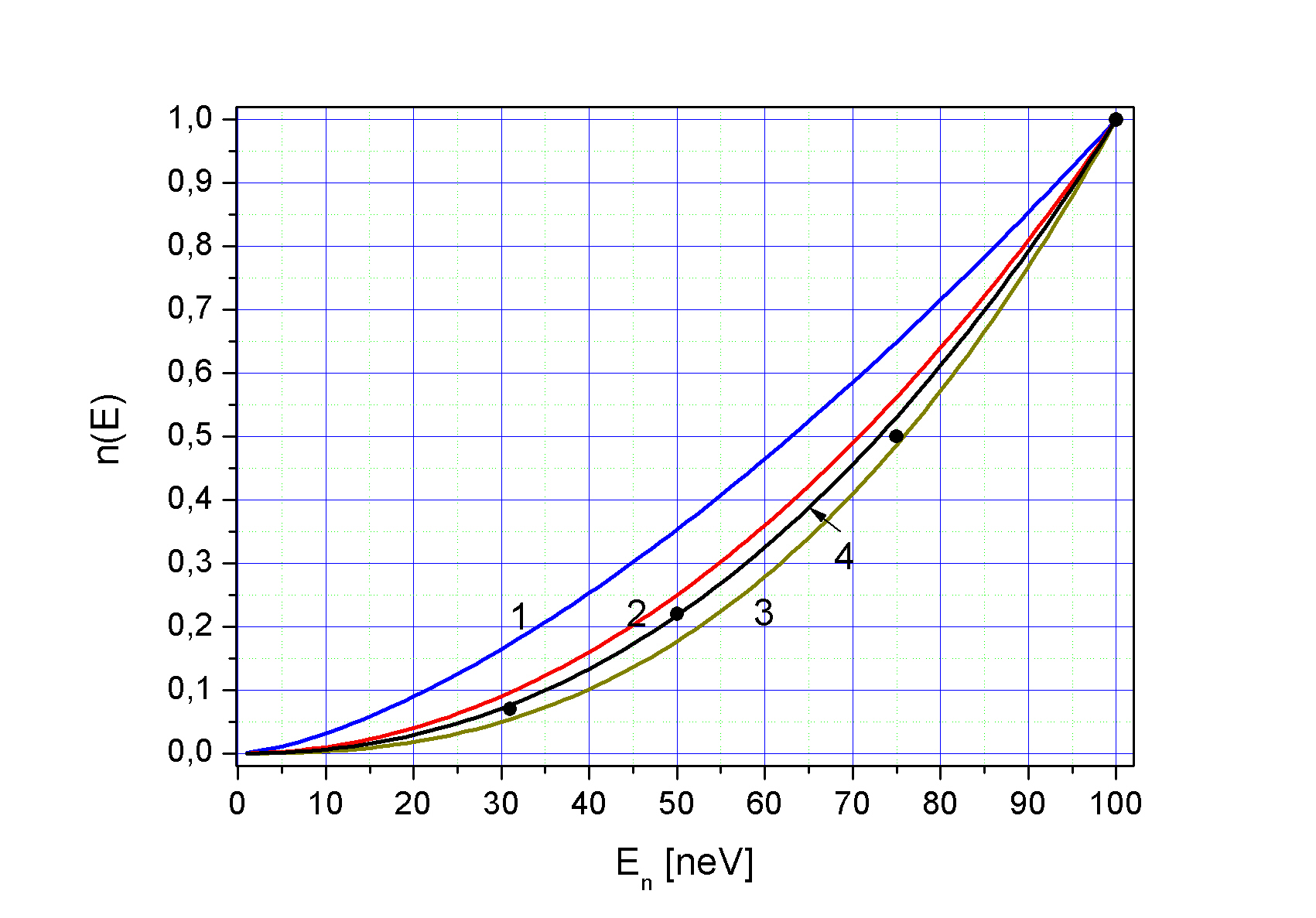}}
\end{center}
\caption{ Dots: numbers of stored neutrons at different heights of the absorber position
extrapolated to zero time and normalized to unity at a height of 100 cm.
 Curves: dependence of the number of stored neutrons on the height of the absorber
 position (also normalized to unity at a height of 100 cm) calculated under different
 assumptions regarding the shape of the initial UCN spectrum:  $n(E)\sim E^{1/2}$ (1),
$n(E)\sim E$ (2), $n(E)\sim E^{3/2}$ (3), $n(E)\sim E^{1.2}$ (4).}
\end{figure}

\newpage
\begin{figure}
\begin{center}
\resizebox{13cm}{13cm}{\includegraphics[width=\columnwidth]{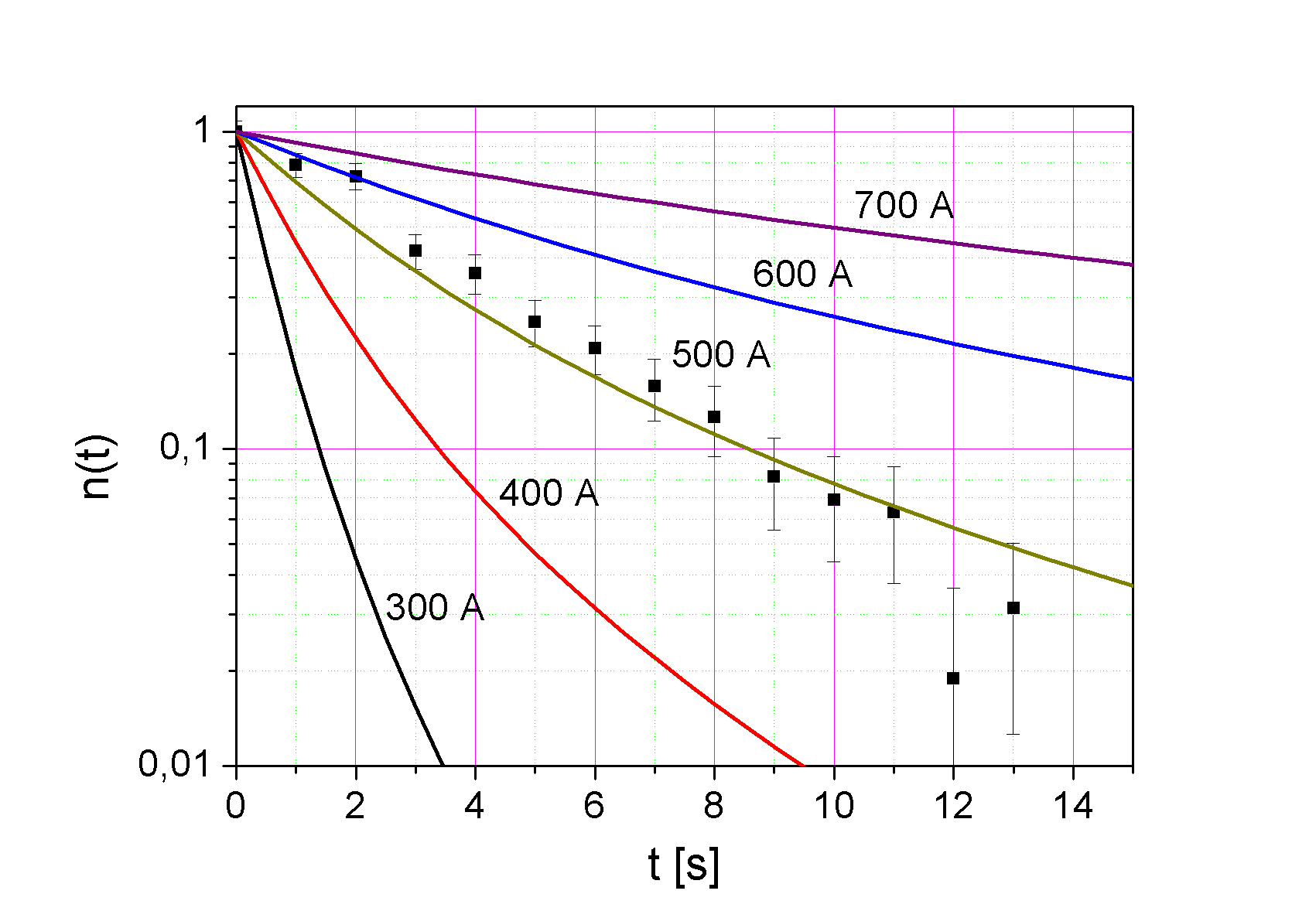}}
\end{center}
\caption{Dots represent the results of measurements of the UCN storage curve in the
titanium chamber.
 Curves were calculated for different thicknesses of the TiO$_{2}$ film.}
\end{figure}

\newpage
\begin{figure}
\begin{center}
\resizebox{13cm}{13cm}{\includegraphics[width=\columnwidth]{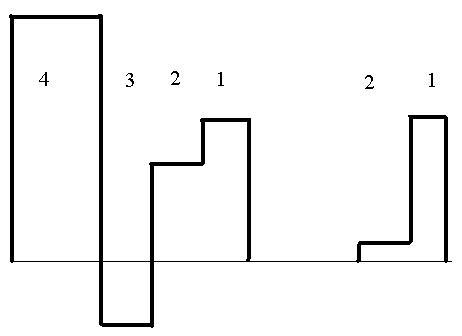}}
\end{center}
\caption{Shapes of potentials used in the UCN reflection calculations.
 Left: UCN storage in the fluoropolymer-coated chamber: (1) fluoropolymer, U = 105 neV;
(2) titanium dioxide TiO${2}$, ReU=68.9 neV, ImU=-1.4$\times 10^{-2}$\,neV; (3)
titanium, ReU=-49.7 neV, ImU=-2.5$\times 10^{-2}$\,neV, copper, Re U=170 neV,
Im U=-2.31$\times 10^{-2}$\,neV.
 Right: calculations of the count rate of a boron counter on the side face of the
 storage chamber: (1) fluoropolymer; (2)  $^{10}B_{0,94}\,^{11}B_{0,06}$,
 $Re U=6,624$\,neV, Im U=-31,4 neV.
 Neutrons are incident from the right.
 The thicknesses are not to scale.}
\end{figure}

\newpage
\begin{figure}
\begin{center}
\resizebox{13cm}{13cm}{\includegraphics[width=\columnwidth]{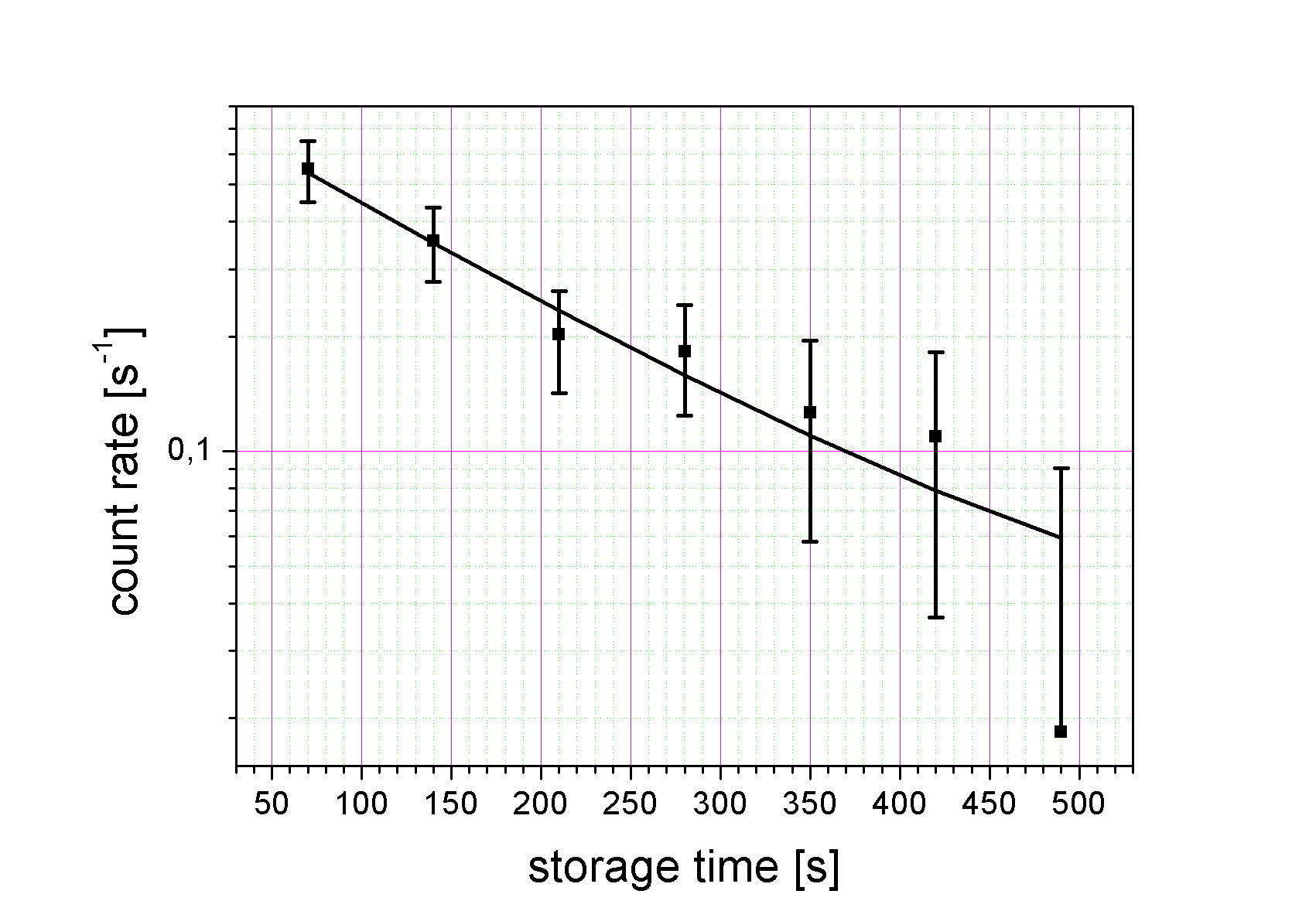}}
\end{center}
\caption{Temporal variation of the count rate of the side boron counter measured in
experiments performed in the fluoropolymer-coated chamber.
 The curve is the result of approximation with a single exponential model.}
\end{figure}

\newpage
\begin{figure}
\begin{center}
\resizebox{13cm}{13cm}{\includegraphics[width=\columnwidth]{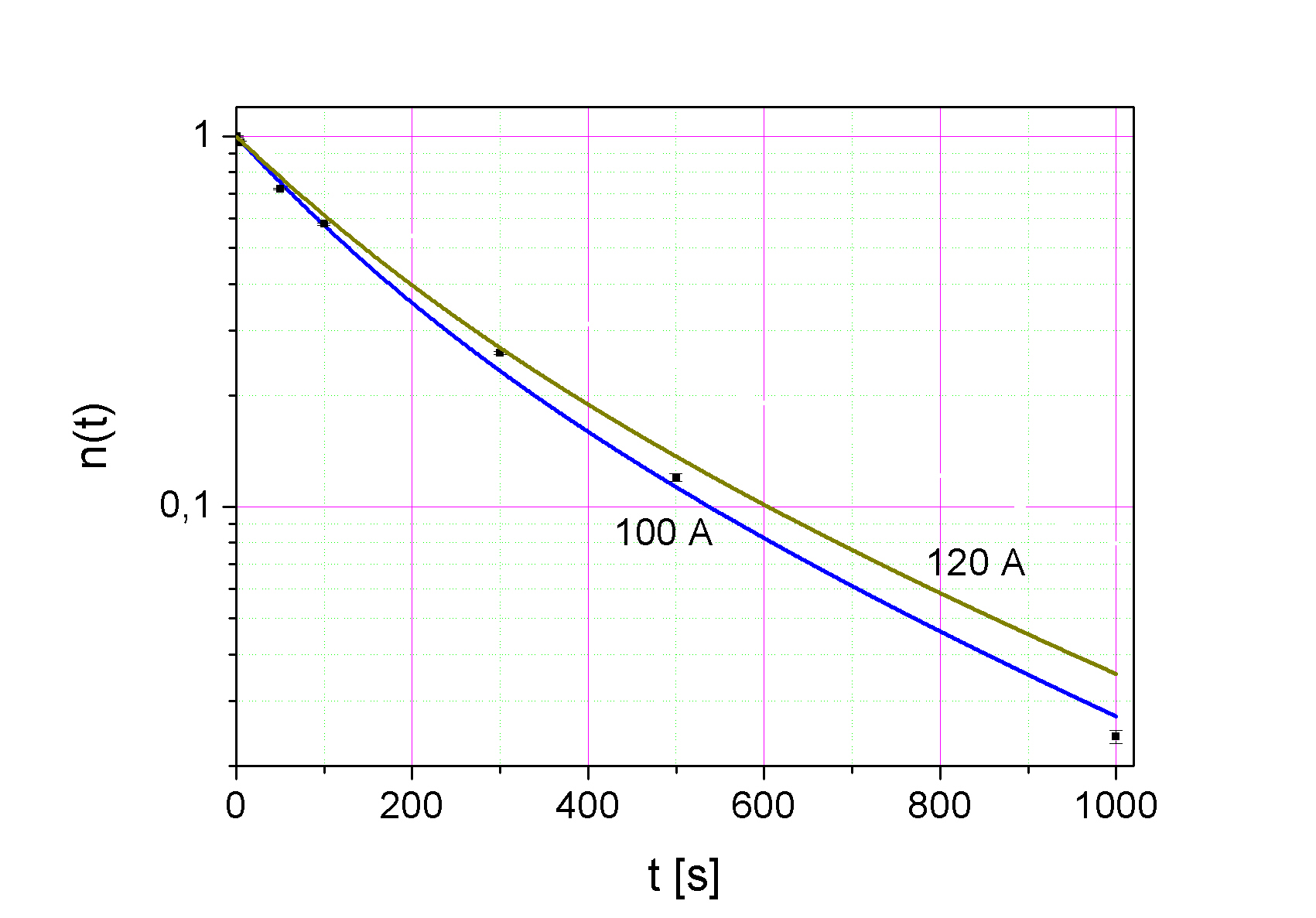}}
\end{center}
\caption{Measured data on the UCN storage in the fluoropolymer-coated chamber
(dots) and storage curves modelled with an assumed fluoropolymer thickness of
650\,\AA and two different thicknesses of the TiO$_{2}$ film (100 and 120\,\AA
).}
\end{figure}
\end{document}